\documentclass{article}
\usepackage{spconf,amsmath,graphicx}
\usepackage{epsf}

\usepackage{epsfig}

\usepackage{epstopdf}
\usepackage{cite}

\newtheorem{definition}{\bf Definition}

\title{A Context-Aware Matching Game for User Association in Wireless Small Cell Networks}

\name {Nima Namvar$^\ast$, Walid Saad$^\dag$, Behrouz Maham$^\ast$, Stefan Valentin$^\ddag$ \thanks{This research was supported by the U.S. National Science Foundation under Grants CNS-$1253731$ and CNS-$1406947$.}}
\address{$^\ast$Electrical and Computer Engineering Department,University of Tehran,Tehran, Iran. \\
	$^\dag$Electrical and Computer Engineering Department, University of Miami, Coral Gables, FL, USA.\\
    $^\ddag$Bell Laboratories, Alcatel-Lucent, Germany.\\
	Emails: {n.namvar@ece.ut.ac.ir, walid@miami.edu, bmaham@ut.ac.ir, stefan.valentin@alcatel-lucent.com}
    }

\begin{document}
\ninept

\maketitle
\begin{abstract}
Small cell networks are seen as a promising technology for boosting the performance of future wireless networks. In this paper, we propose a novel context-aware user-cell association approach for small cell networks that exploits the information about the velocity and trajectory of the users while also taking into account their quality of service (QoS) requirements. We formulate the problem in the framework of matching theory with externalities in which the agents, namely users and small cell base stations (SCBSs), have strict interdependent preferences over the members of the opposite set. To solve the problem, we propose a novel algorithm that leads to a stable matching among the users and SCBSs. We show that the proposed approach can better balance the traffic among the cells while also satisfying the QoS of the users. Simulation results show that the proposed matching  algorithm yields significant performance advantages relative to traditional context-unaware approaches.
\end{abstract}

\begin{keywords}
Small cell network; Context information; User-cell association; Matching theory.
\end{keywords}

\section{Introduction}\vspace{-0.1cm}\label{SEC:Intro}

Present-day user equipments (UEs) and accompanying wireless services require significant wireless resources and often impose highly fluctuating traffic into cellular networks \cite{cisco,3GPP,JSAC Femto}. Recently, considerable attention has been devoted to the concept of small cell networks (SCNs) as a suitable and promising approach to cope with this unprecedented increase in wireless data traffic \cite{Guvencbook}. While the deployment of small cell networks carries considerable potential benefits, it requires meeting several new technical challenges such as interference management, network planning, and self-organization \cite{JAndrewsJournal}.

In particular, one fundamental technical challenge in SCNs is that of user association \cite{KtierHetNet}. Indeed, SCNs consist of different types of access points (femtocells, picocells, microcells) with different size, capacity, and capabilities. Hence, due to this heterogeneous nature of SCNs, applying traditional macro-cell oriented user-cell association (UCA) strategies to SCNs, may yield undesirable outcomes \cite{JAndrewsJournal2}. Thus, developing and designing new UCA strategies that are tailored to the specific nature of wireless SCNs is needed.

In \cite{JAndrewsJournal2}, a user association strategy is studied in SCNs which aims at balancing the load in SCNs by optimizing overall long-term rate. An adaptive UCA approach for the downlink of SCNs is proposed in \cite{FlexibleAssociation}. This approach exploits different biasing factors in the received signal to noise ratio (SNR) of different tiers to fairly distribute the traffic among the cells. Simple UCA approaches compatible with the Long Term Evolution-Advanced (LTE-A), such as range expansion and cell-splitting are presented in \cite{JSACUCA}. The work in \cite{decentralized} proposes a decentralized UCA method for the downlink SCNs, which is based on an iterative scheme that exploits the feedback information of each individual user to enhance the network's quality of service (QoS). Other related works are found in \cite{MACA,DLB,MANET}. These works focused on UCA approaches that rely on limited, conventional network information such as signal strength or channel state. However, by utilizing readily available additional information about different layers of network, such as the users' mobility pattern and the urgency of traffic each user imposes to the network, more efficient approaches for user association could be introduced, as proposed in this paper.

Here, we refer to such additional information as the users' \emph{context} information. The works in  \cite{JAndrewsJournal2,FlexibleAssociation,JSACUCA,decentralized,MACA} do not exploit context information; although the use of such context can improve the overall UCA performance. For instance, by knowing the traffic urgency of different users that are requesting network's resources, the network could better address the traffic by prioritizing the more urgent users and postponing the service to the less urgent ones \cite{ValentinConf}. We will show that by using such suitable context information, the network can better decide that which user should be serviced by which SCBS while also satisfying the QoS requirements of the users.

The main contribution of this paper is to develop a novel UCA scheme in the downlink of SCNs by exploiting a combination of new context information related to the users trajectory profile, their QoS demands, and the current load of each cell. The effects of context information on the UCA strategy are effectively modeled via well-designed utility functions. We formulate the UCA problem as a many-to-one matching game with externalities, in which the preferences of the players, i.e. users and SCBSs, are interdependent. Such interdependency which stems from the mutual interference, significantly changes the game properties as opposed to classical matching games \cite{Leshem}, \cite{Jorswieck}. To solve the proposed matching game, we propose a novel self-organizing algorithm that can dynamically update the preference lists in the presence of externalities and is able to reach to a stable matching between the users and their serving SCBSs. Simulation results assess the performance of the proposed approach and show that it yields noticeable performance gains relative to context-unaware approaches. To our best knowledge, although some works such as and \cite{ValentinConf} and \cite{ValentinJournal} have explored application information for scheduling in conventional cellular networks, the combination of these three context information to optimize the problem of UCA in SCNs \emph{has not been} investigated in the prior-art.

The rest of this paper is organized as follows: The system model is presented in Section \ref{SEC:SysModel}. In Section \ref{SEC:Matching}, the problem of user-SCBS association is studied in the framework of matching theory with externalities and a novel algorithm for solving the game is proposed. Simulation results are provided in Section \ref{SEC:Simulation} and finally, conclusions are drawn in Section \ref{SEC:Conlusion}.\vspace{-0.3cm}

\section{System Model}\vspace{-0.1cm}\label{SEC:SysModel}
Consider the downlink of a two-tier wireless small cell network consisting of macrocells and picocells. Let $\mathcal{M}$, $\mathcal{P}$ and $\mathcal{N}$ denote the set of $\textit{M}$ MBSs, the set of $\textit{P}$ pico base stations (PBSs), and the set of $\textit{N}$ users respectively. The wireless channel suffers form multipath slow fading. Each user follows a certain mobility path through the network and possesses its own, specific QoS demand. For each user, we associate a unique profile defined by three variables $\tau, \theta, V$ representing the urgency of the service in use, the direction of its motion, and its velocity, respectively. Next, we study these parameters and we show how such information can improve the UCA strategies. \vspace{-0.3cm}

\subsection{QoS requirement}
Depending on their traffic patterns and service requirements, the users may have varying needs for the wireless resources. For example, a user engaged in live video streaming requires a high data rate and its QoS vitally depends on the delivery time. In contrast, a file-downloading application is not too susceptible to delay. Context-unaware UCA approaches, do not prioritize the users based on their QoS needs \cite{ValentinJournal}; however, as we will see, considering the data urgency could help maintaining higher average QoS for the users.

The quality of experience (QoE) for each user can be defined as a decreasing function of delivery time. The abruptness of QoE drop over time depends on the data urgency, i.e. the more urgent data, the more rapid fall of QoE as time elapses. Some suggestions to quantitatively model such behavior are presented in \cite{ValentinConf} and \cite{ValentinJournal}. We assume that the quality of experience (QoE) of each user $n\in \mathcal{N}$, varies with the delivery time as follows:

\begin{equation}
Q_n(t) = \frac{1}{1+\textrm{exp}(t-\tau_n)},
\end{equation}
where $\tau_n$ is the urgency coefficient, which accounts for the urgency of the data. The smaller is $\tau_n$, the more urgent is the data in use. Indeed, this function effectively captures the QoE variations over time, as described before. \vspace{-0.3cm}

\begin{figure}\label{FIG:coverage}
  \begin{center}
   \vspace{-0.2cm}
    \includegraphics[width=9cm]{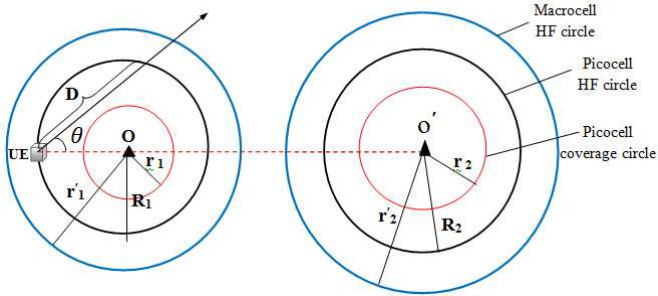}
    \vspace{-0.8cm}
    \caption{The HF and coverage region}\vspace{-0.8cm}
  \end{center}
\end{figure}

\subsection{Trajectory and handover failure probability}\vspace{-0.1cm}
Users travel through the network and active communication sessions must be transferred among the cells. Thus, the users' mobility can lead to a number of  handover (HO) processes within the SCN. To guarantee the QoS, the network should avoid risky HOs that may lead to a signal loss or erroneous communication. A handover fails when the received SINR drops under a certain threshold \cite{GuvencHandover}.

Without loss of generality, we assume circular coverage area for tractability. Fig. \ref{FIG:coverage} shows the handover scenario under consideration. We consider two picocells in the vicinity of each other and assume that a user is traveling through these two cells. Here, we study the probability of handover failure (HF) considering some context information such as the users' speed and trajectory. When a users enters a cell, the total possible time of interaction $t_{T}$ could be computed as:
\begin{equation}\label{Intraction time}
  t_{T}=\frac{D}{V},
\end{equation}
where $D$, as shown in Fig. 1, represents the distance covered by a user when passing through the coverage area of the cell and $V$ is its average speed. Hereinafter, we assume that $V$ is small and that the users have \emph{low} to \emph{medium} mobility. A successful HO process needs a certain preparation time of duration $T_p$ before it could be initiated. Thus, based on the values of $t_T$ and $T_p$, there are two different scenarios: If $t_T > T_p$, the user is considered as a \emph{candidate} to be served; otherwise, no HO would be initiated.

Here, $D=2R\textrm{cos($\theta$)}$ represents the chord of the coverage circle, where $R$ is the radius of the coverage circle and $\theta$ is considered to be the angle of trajectory. Since the directions which the user takes into the picocell are equally likely, $\theta$ is a random variable uniformly distributed between $-90^{\circ}$ and $+90^{\circ}$. Therefore, the cumulative distribution function (CDF) of $D$ could be derived as: \vspace{-0.5cm}

\begin{equation}
 \textrm{Pr}(D<d)= 2\textrm{Pr}(\theta>\textrm{cos}^{-1}\left(\frac{d}{2R}\right))=1-\frac{2}{\pi}\textrm{cos}^{-1}\left(\frac{d}{2R}\right).
\end{equation}
When the path is the tangent of the HF circle (with the radius $r$), $D$ is equal to $2\sqrt{R^2-r^2}$. HF occurs when the MUE's path intersects the circle of HF failure, thus the probability of HF is:
\begin{equation}
  \textrm{Pr} (HF) = \textrm{Pr}(D\geq2\sqrt{R^2-r^2})=\frac{2}{\pi}\textrm{cos}^{-1}(\sqrt{1-(\frac{r}{R})^2}).
\end{equation}
From (3), it can be seen that the HO process becomes more reliable as $r$ becomes smaller relatively to $R$. The ratio of $r$ to $R$ varies for each cell and so, the different cells guarantee different levels of reliability in HO process.

Now, assume that a user exits from the picocell $p_1$ and enters another picocell $p_2$. A successful handover could be initiated as soon as the user leaves $p_1$ and must be terminated before the user's distance from $p_1$ exceeds $r'_1>R_1$ and also before it runs into the $p_2$ to the level of $r_2$. Let $O$ and $O'$ represent the centers of $p_1$ and $p_2$ respectively, then $OO'$ is the distance between the two PBSs. So, the following conditions determine the possible initiation of HO: \newline
1) If $OO'> r'_1+R_2$ or $OO'<R_1+r_2$ : No HO would be initiated \newline
2) If $R_1+r_2 \leq OO'\leq r'_1+R_2$, then a HO could be initiated.

 The probability of error in in the second case can be given by:
\begin{multline}\
  \textrm{Pr}(HF)=1-\textrm{Pr}(\text{successful HO}) \\
  =1-\textrm{Pr}(V<\frac{r'_1-R_1}{t_{m_1}})\times (1-\frac{2}{\pi}\textrm{cos}^{-1}(\sqrt{1-(\frac{r_2}{R_2})^2}))
\end{multline}

Given the defined context information, in the next section, we formulate the UCA problem as a one-to-many matching game.

\section{Cell Association as a matching game with externalities}\vspace{-0.1cm}\label{SEC:Matching}
Context aware UCA problem has a striking analogy to the problem of college admission studied in matching theory literature such as \cite{Roth}, \cite{Salgado}. Here, we formulate the UCA problem in the matching theory framework. Consider the set $\mathcal{N}=\{n_1,n_2,...,n_N\}$ of all $N$ users and let $\mathcal{P}=\{p_1,p_2,...p_P\}$ be the set of $P$ SCBSs. The outcome of the UCA problem is a \textit{matching} between two sets $\mathcal{N}$ and $\mathcal{P}$ which is defined as follows:

\begin{definition} \hspace{-0.1cm}
A matching $\mathcal{\mu}$ is a function from $\mathcal{N}\cup \mathcal{P}$ to $2^{\mathcal{N}\cup \mathcal{P}}$ such that $\forall n\in \mathcal{N}$ and $\forall p\in \mathcal{P}$: (i) $\mu(n)\in \mathcal{P}\cup \emptyset$ and $|\mu(n)|\leq1$, (ii) $\mu(p)\in 2^{\mathcal{N}}$ and $|\mu(p)|\leq q_p$ where $q_p$ is the quota of $p$, and (iii) $\mu(n)=p$ if and only if $n$ is in $\mu(p)$.
\end{definition}
The users who are not assigned to any member of $\mathcal{P}$, will be assigned to the nearest MBS. Members of $\mathcal{N}$ and $\mathcal{P}$ must have strict, reflexive and transitive preferences over the agents in the opposite set \cite{Salgado}. In the next subsections, exploiting the context information we introduce some properly-defined utility functions (UFs) to effectively capture the preferences of each set.

\subsection{Users' preferences}\vspace{0.1cm}
Users demand for reliable and high quality communication. Therefore, they prefer the SCBSs which can guarantee the safety of communication during the handover and are able to deliver the data in an acceptable time.

To guarantee a reliable communication, the HO could be initiated only if the HF probability is less than an acceptable threshold. For example, let the threshold be $\textrm{Pr}(HF)<5\%$, then the next cell must hold this condition: $\frac{r_m}{R}\leq 0.08$. If the cell does not satisfy this condition, then, it would be \textit{unacceptable} to the user. When the HF probability is small enough, (3) could be simplified using Maclaurin series expansion to a linear relationship, i.e.  $Pr(HF)\simeq \frac{2 r_m}{\pi R}$, which can be used directly as metric of the handover reliability.

Moreover, the users seek to optimize their transmission rate which depends on the current cell load and the interference caused by the neighbor SCNs. Therefore, the rate of transmission is a function of current matching $\mu$. We define the rate over load for all user-SCBS pairs as:
\begin{equation}
  \mathcal{r}_{ij}=\frac{1}{\textrm{max}(1,K_j)} \textrm{log}_2(1+\frac{P_jc_{ij}}{\sum_{k \neq j} P_kc_{ik}+\sigma^2}),
\end{equation}
Where $K_j$ is the total users being served by SCBS $j$, $P_j$ is the power of SCBS $j$, and $c_{ij}$ represents the channel coefficient between user $i$ and SCBS $j$. $\sigma^2$ is the power of additive noise.

We define the following utility that user $i$ gains if admitted by SCBS $j$:
\begin{equation}
 U_{i}\left(\mu,R_j,r_j,\mathcal{r}_{ij}\right)=\frac{R_j}{r_j}\mathcal{r}_{ij} \hspace{3mm},
\end{equation}
It is seen that the users prefer lightly-loaded SCBSs to maximize their utility. This utility function could help to offload the heavily-loaded cells by pushing the users to more lightly-loaded cells. (7) also captures the user's natural objective to maximize its transmission rate. Using (7), the users can rank the SCBSs in their vicinity based on the defined utility.

\subsection{Small cells' preferences}\vspace{-0.1cm}
The main goal of SCBSs is to increase the network-wide capacity by offloading traffic from the MBSs while providing satisfactory QoS for the users. Each SCBS $p$ could serve a limited number of users, called quota $q_p$. Assume that user $n$ enters the coverage area of small cell $p$ and requests to be serviced by $p$. The cell would only consider the users who meet the following condition:
\begin{equation}\
  \text{$n\in \mathcal{N}$ is acceptable to $p\in \mathcal{P}$} \Leftrightarrow \frac{D_n}{V_n}\geq T_p,
\end{equation}
where $T_p$ represents the required preparation time of SCBS $p$.

By prioritizing the users coming from congested cells, the SCBSs could offload the heavily-loaded cells. On the one hand, each candidate user is carrying a potential utility as a function of the pervious cell $p'$' load, i.e. $f(\frac{K_{p'}}{q_{p'}})$. This utility depends on the current matching which determines the number of users in neighboring cells. On the other hand, SCBS $p$ would also prioritize the candidate users based on their QoS guarantee requirement defined in (1). We define the following utility that each SCBS $p$ obtains by serving an acceptable UE $n$:
\begin{equation}
V_{p}\left(\mu,\tau_n,k_{p'},q_{p'}\right)= \left[1+\textrm{log}\left(\frac{\textrm{max}(1,k_{p'})}{q_{p'}}\right)\right] \frac{1}{\tau_n}.
\end{equation}

The first term in (9) accounts for the offloading concept, and the second term is the utility achieved by the SCBS $p$ for its service to a specific application. Indeed, the small cell $p$ gains more utility by giving service to the users having more urgent data.

From (7) and (9), we can see that the utilities depend on the current matching $\mu$ and consequently, the preferences of agents are interdependent. Therefore, the preferences of players are not solely based on individuals, but some \emph{externalities} affect the preferences and matching as well \cite{ExternalBando}. Let $\Psi(\mathcal{N},\mathcal{P})$ be the set of matchings.

\begin{definition}
The preference relation $\succ_n$ of the user $n\in \mathcal{N}$ over the set $\Psi(\mathcal{N},\mathcal{P})$ is a function that compare two matchings $\mu, \mu' \in \Psi$ such that:
\begin{equation}
  \mu \succ_n \mu' \Leftrightarrow U_n(\mu)>U_n (\mu').
\end{equation}
\end{definition}
The preference relation for the SCBSs $\succ_p$ is defined similarly. Users and SCBSs rank the members of the opposite set based on the defined preference relations. Our purpose is to match the users to the small cells so that the preferences of both side be satisfied as much as possible; thereby the network-wide efficiency would be optimized.

To solve a matching game, one suitable concept is that of a stable matching. In a matching game with externalities, stability has different definitions based on the application. Here, we consider the following notion of stability in the problems with externalities \cite{Salgado}:

\begin{definition}
A matching $\mu$  is blocked by coalition of the users $n$ and the SCBS $p$ if there exists $\mu' \in \Psi$ so that $\mu'(n) \succ_n \mu(n)$ and $\mu'(p) \succ_p \mu(p)$. A many to one matching is \emph{stable} if it is not blocked by any coalition of users and SCBSs.
\end{definition}

In the next section, an efficient algorithm for solving the game is presented, which reaches to a stable matching between users and small cells.

 \begin{table}[t] \vspace{-0.3cm}
  \scriptsize
  \centering
  \caption{Proposed Algorithm For The User-Cell Association Game}
    \begin{tabular}{p{8.5cm}}
      \hline

\textbf{Input:} context-aware utilities and the preferences of each user \\
\textbf{Output:} Stable matching between the D2D pairs \\
\\
\textbf{Initializing}: All the UEs are assigned to the nearest BS \\
\\
\textbf{Stage I}: \textbf{Preference Lists Composition}
\begin{itemize}
  \item Neighboring D2D users exchange their context information
  \item Users sort the set of acceptable candidate based on their preference functions
\end{itemize} \\
\textbf{Stage II}: \textbf{Matching Evaluation} \\
\hspace*{1em}\textbf{while:} $\mu^{(n+1)}\neq \mu^{(n)}$
\begin{itemize}
  \item Update the utilities based on the current matching $\mu$
  \item Construct the preference lists using preference relations
  \item Each user $n$ applies to its most preferred partner
  \item Each user accept the most preferred applicants and create a waiting list while rejecting the others
\end{itemize}
\hspace*{2em} \textbf{Repeat}\\
\hspace*{4em} $\bullet$ Each rejected user applies to its next preferred partner \\
\hspace*{4em} $\bullet$ Each user update its waiting list considering the new applicants\\
\hspace*{2em} \textbf{Until:} all the users assigned to a waiting list \\
\hspace*{1em}\textbf{end} \\
   \hline
    \end{tabular}\label{tab:algo}\vspace{-0.6cm}

\end{table}

\subsection{Proposed Algorithm}\vspace{-0.1cm}
The deferred acceptance algorithm, introduced in \cite{Roth}, is a well-known approach to solving the standard matching games. However, in our game, the preferences of agents as shown in (7) and (9), depend on externalities through the entire matching, unlike classical matching problems in which preferences are static and independent of the matching. Therefore, the classical approaches such as the deferred acceptance cannot be used here because of the presence of externalities \cite{Salgado},\cite{ExternalBando}. To solve the formulated game, we propose a novel algorithm shown in Table I. Suppose that all the users are initially associated to the nearest MBS. Each user sends its profile information ($V$, $\alpha$, $\tau$) to the neighboring SCBSs. Each SCBS, on the other side, only keeps the users satisfying (8) and ranks them based on their utilities (9). Upon ranking the acceptable UEs, the SCBS feeds back the awaiting users with its own context information including its rate over load defined in (6) and its corresponding coverage and HF circle radii $R$ and $r$.

Each user makes a ranking list of the available SCBSs and applies to the most preferred one of them. The SCBSs rank the applicants and keep the most preferred ones up to their quota and reject the others. The users who have been rejected in the former phase, would apply to their next favorite SCBS and the SCBSs modify their waiting list accordingly. This procedure continues until all the users assigned to a waiting list.

 However, since the preferences depend on the current matching $\mu$, an iterative approach should be employed. In each step, the utilities would be updated based on the current matching. Once the utilities are updated, the preference lists would be updated accordingly as well. Therefore, in each iteration, a new temporal matching arises and based on this matching, the interdependent utilities are updated as well. The algorithm initiates the next iteration based on the modified preferences. The iterations run on until two subsequent temporal matchings are the same and algorithm converges.

The proposed algorithm will lead to a stable matching when it converges. Indeed, the deferred acceptance in stage II would not converge if the matching is not stable \cite{Roth}. Hence, by contradiction, whenever the algorithm converges, the matching would be stable.

\begin{figure}\vspace{-0.2em}
  \begin{center}
   \vspace{-0.2cm}
    \includegraphics[width=7.9cm]{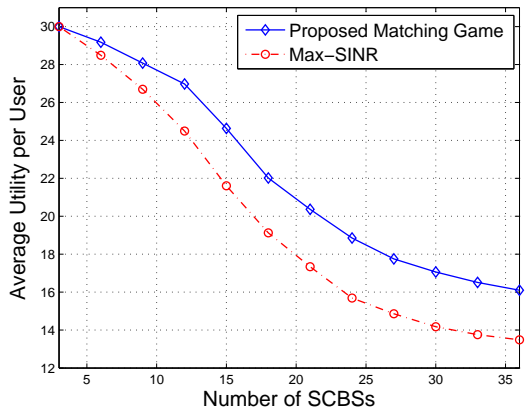}
    \vspace{-0.5cm}
    \caption{Average utility per user for different number of SCBSs with $N=60$ users.}\vspace{-2em}
  \end{center}
\end{figure}\vspace{-0.3em}

\section{Simulation Results}\vspace{-0.1cm}\label{SEC:Simulation}
For our simulations, we consider a single MBS with radius 1 km and  overlaid by $P$ uniformly deployed picocells. The small cells' quota is supposed to be a typical value $q=$4 for all SCBSs \cite{StandardFemto}. The channels suffer a Rayleigh fading with parameter $\sigma=2$. Noise level is assumed to be $\sigma^2=-121$~dBm and the minimum acceptable SINR for the UEs is 9.56 dB \cite{Sesia}. There are $N$ users distributed uniformly in the network. The QoS parameter $\tau_n$ in (1) is chosen randomly from the interval [0.5,5] ms. The users have low mobility and can be assumed approximately static during the process time required for a matching. All the statistical results are averaged by 1000+ runs over random location of users and SCBSs, the channel fading coefficients, and other random parameters.

The performance is compared with the max-SINR algorithm which is a well-known context-unaware approach exploited in wireless cellular networks for the UCA. In this approach, each user is associated to the SCBS providing the strongest SINR.
\begin{figure}\vspace{1em}
  \begin{center}
   \vspace{-0.2cm}
    \includegraphics[width=7.9cm]{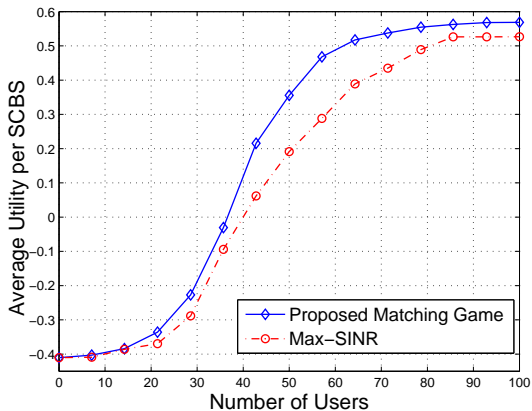}
    \vspace{-0.5cm}
    \caption{Average utility per SCBS for different number of users with $P=20$ SCBSs.}\vspace{-2em}
  \end{center}
\end{figure}\vspace{-0.1em}

Fig. 2 shows the average utility per user as a function of the number of SCBSs for $N=60$ users. As the number of SCBSs increases, the average load in each cell decreases. Consequently, the available resources for each user would increase. However, increasing the number of SCBSs leads to an increase in the interference between the neighboring cells and, consequently decreases the rate. Hence, as shown Fig. 2, the average utility per user is a decreasing function with respect to the number of SCBSs. We can also see that, when the number of SCBSs exceeds $30$, the slope of the curves becomes less steep due to the fact that the matchings would not be affected by these increments as much as before. Indeed, beyond this point, there are enough SCBSs to serve all the users and deploying more SCBSs will not change the current matching dramatically. Fig. 2 demonstrates that the proposed context-aware approach has a considerable performance advantage compared to the max-SINR approach. This performance advantage reaches up to $20.4\%$ gain over to max-SINR criterion for a network with 30 SCBSs.

Fig. 3 shows the average utility achieved by each SCBS as a function of the number of users for $P=20$ SCBSs. As the number of users $N$ increases, the network becomes more congested, and the probability that a new user who applies for an SCBS is coming from a congested BS increases. Therefore, it is more likely for the SCBSs to gain more utility by offloading the network. However, when the network is considerably congested, the new users that arrive to the network would be mostly assigned to the MBS, since many of SCBSs are already servicing their maximum capacity. In this respect, Fig. 3 shows that, once the number of users exceeds $80$, the average utility of SCBSs remains constant. The proposed algorithm achieves up to 24.9\% gain over the Max-SINR approach when the number of users is 50.

Fig. 4 shows the average number of required iterations per user required for the algorithm to converge to a stable matching for two different network sizes, as the number of users varies. In this figure, we can see that the number of algorithm iterations is an increasing function of the number of users and the number of SCBSs. Fig. 4 shows that the average number of iterations varies from $1.09$ and $1.1$ at $N=3$ to $5$ and $8.6$ at $N=70$, for the cases of $10$~SCBSs and $20$ SCBSs, respectively. Clearly, Fig. 4 demonstrates that the proposed algorithm converges in a reasonable number of iterations.

\begin{figure}\vspace{-0.6em}
  \begin{center}
   \vspace{-0.2cm}
    \includegraphics[width=7.9cm]{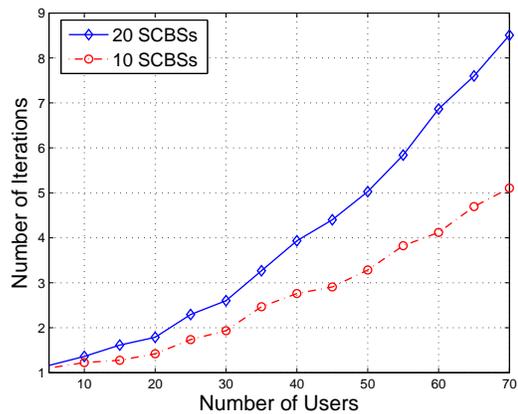}
    \vspace{-0.5cm}
    \caption{Average number of iterations per user to reach a stable matching for different values of SCBSs.}\vspace{-3em}
  \end{center}
\end{figure}\vspace{-0.1em}

\vspace{-1.1em}

\section{Conclusions}\vspace{-0.2cm}\label{SEC:Conlusion}
In this paper, we have proposed a new context-aware user association algorithm for the downlink of the small cell networks. By introducing well-designed utility functions, our approach accounts for the trajectory and speed of the users as well as for their heterogeneous QoS requirements. We have modeled the problem as a many-to-one matching game with externalities, where the preferences of the players are interdependent and contingent on the current matching. To solve the game, we have proposed a novel algorithm that converges to a stable matching in a reasonable number of iterations. Simulation results have shown that the proposed approach yields considerable gains compared to max-SINR approach.

\bibliographystyle{ieeetr}
\bibliography{myref}

\end{document}